\documentclass[preprint,aps,showpacs]{revtex4}
\usepackage{mathrsfs}
\usepackage{amsmath}
\usepackage{amssymb}
\usepackage{epsfig}
\usepackage{graphicx}
\usepackage{booktabs}
\usepackage{array}
\usepackage{multirow}
\begin{document}
\renewcommand{\arraystretch}{0.5}
\newcommand{\beq}{\begin{eqnarray}}
\newcommand{\eeq}{\end{eqnarray}}
\newcommand{\non}{\nonumber\\ }

\newcommand{\acp}{ {\cal A}_{CP} }
\newcommand{\psl}{ p \hspace{-1.8truemm}/ }
\newcommand{\nsl}{ n \hspace{-2.2truemm}/ }
\newcommand{\vsl}{ v \hspace{-2.2truemm}/ }
\newcommand{\epsl}{\epsilon \hspace{-1.8truemm}/\,  }

\def \cpl{ Chin. Phys. Lett.  }
\def \ctp{ Commun. Theor. Phys.  }
\def \epjc{ Eur. Phys. J. C }
\def \jpg{  J. Phys. G }
\def \npb{  Nucl. Phys. B }
\def \plb{  Phys. Lett. B }
\def \prd{  Phys. Rev. D }
\def \prl{  Phys. Rev. Lett.  }
\def \zpc{  Z. Phys. C }
\def \jhep{ J. High Energy Phys.  }

\title{The pure annihilation type decays $B^0\rightarrow D_{s}^{-}K_{2}^{*+}$ and $B_{s}\rightarrow \bar{D}a_{2}$ in perturbative QCD approach}
\author{Zhi-Tian Zou}
\author{Zhou Rui}
\author{Cai-Dian L\"{u}}\email{lucd@ihep.ac.cn}
\affiliation{Institute  of  High  Energy  Physics  and  Theoretical  Physics Center for Science Facilities,
Chinese Academy of Sciences, Beijing 100049, People's Republic of China }

\date{\today}
\begin{abstract}
We calculate the branching ratios of pure annihilation type decays
$B^0\rightarrow D_{s}^{-}K_{2}^{*+}$ and $B_s\rightarrow
\bar{D}a_{2}$ using the perturbative QCD approach based on $k_T$
factorization. The branching ratios are predicted to be
$(60.6_{-16.5\,-10.4\,-2.1}^{+17.3\,+4.3\,+3.2})\times 10^{-6}$ for
$B^0\rightarrow D_{s}^{-}K_{2}^{*+}$,
 $(1.1_{-0.4\,-0.2\,-0.1}^{+0.4\,+0.1\,+0.1})\times 10^{-6}$ for $B_s\rightarrow \bar{D}a_{2}^{0}$ and $(2.3_{-0.8\,-0.4\,-0.1}^{+0.8\,+0.2\,+0.1})\times 10^{-6}$ for $B_{s}\rightarrow D^{-}a_{2}^{+}$.
 They are large enough to be measured in the ongoing experiment. Due to the shortage of contributions from penguin operators,
   there are no direct CP asymmetries for these decays in the standard
   model. We also derive simple relations among these decay channels to reduce theoretical uncertainties for the
   experiments to test the accuracy of theory and search of new physics signal.
\end{abstract}



\maketitle

\section{Introduction}

 Two body hadronic  B decays have been a hot topic for many years,
since it involves the perturbative QCD calculation and factorization
study. It is also important for the test of standard model, the CKM
angle measurements and the search of new physics phenomena. For many
years, people do calculations based on the naive factorization
assumption, later proved by the soft-collinear effective
theory\cite{scet}. However, there is one kind of diagrams, the so
called annihilation type diagrams, which was argued to be helicity
suppressed since no one  knows how to calculate. In the well
developed collinear factorization, there is endpoint singularity in
the calculation of these diagrams. In fact, this kind of diagrams
are essential for the strong phase of direct CP asymmetry in the
$B\to K^+\pi^-$ decays \cite{directcp}, which is proved to be
important.

Furthermore, there is one kind of B decays, which contains only
annihilation type diagram contributions. One of the examples is the
$B^0 \to D_s K^+$ decay, which is predicted in
ref.\cite{dsk,prd78014018,jpg37015002} and measured by the B
factories later \cite{exp}. Recently, the CDF collaboration measured the
first pure annihilation type decays in the $B_s$ sector i.e. $B_s
\to \pi^+ \pi^-$ decay, which exactly confirms the perturbative QCD
prediction for this decay \cite{annihilation1,prd76074018}. It is worth of mentioning
that the perturbative QCD (PQCD) approach is almost the only method
can do the quantitative calculations of the  annihilation type
diagrams \cite{annihilation1,prd76074018}.

In this paper, we shall study the pure annihilation type charmed
 decays $B^0\rightarrow D_{s}^{-}K_{2}^{*+}$ and $B_s\rightarrow \bar{D}a_{2}$ in the PQCD approach, which
is based on the $k_{T}$ factorization \cite{wang7,prd63074009}.
These decays are predicted to have a large branching ratio as
$10^{-6}$ to $10^{-5}$, which are measurable in the near future
experiments. In the annihilation type diagrams, both of the light
quark and the heavy anti b quark in B meson annihilate into another
quark anti-quark pair through the four quark operators, while
another light quark pair  in the final state mesons are produce by a
gluon attaching to the four quark operator. Since the light quark in
the final states are  collinear, the  gluon connection them must be
hard. So the hard part of the PQCD approach contains six quarks
rather than four quarks. This is called six-quark effective theory
or six-quark operator. In this approach, the quarks' intrinsic
transverse momenta are kept to avoid the endpoint divergence.
Because of the additional energy scale introduced by the transverse
momentum, double logarithms will appear in the QCD radiative
corrections. We resum these double logarithms to give a Sudakov
factor, which effectively suppresses the end-point region
contribution. This makes the PQCD approach more reliable and
consistent.

This paper is organized as following. In Sec.II, we present the
formalism and perform the perturbative calculations for considered decay
channels with the PQCD approach. The numerical results and
phenomenological analysis are given in Sec.III. Finally, Sec.IV
contains a short summary.

\section{FORMALISM AND Perturbative calculation}\label{sec:function}

The $B^0\rightarrow D_s^-K_2^{*+}$,$B_s\rightarrow
\bar{D}^0a_2^0$ and $B_s\rightarrow
D^-a_2^+$ decays are pure annihilation type rare decays. At the quark level, these decays are described by the effective Hamiltonian
$H_{eff}$ \cite{rmp68}
\begin{eqnarray}
H_{eff}=\frac{G_{F}}{\sqrt{2}}\,V_{cb}^{*}V_{uD}\left[C_{1}(\mu)O_{1}(\mu)\,+\,C_{2}(\mu)O_{2}(\mu)\right],
\label{HH}
\end{eqnarray}
where $V_{cb}$ and $V_{uD}$ are CKM matrix elements, $``D"$ denotes
the light down quark d or s, and $C_{1,2}(\mu)$ are Wilson
coefficients at the renormalization scale $\mu$. $O_{1,2}(\mu)$ are
the four quark operators.
\begin{eqnarray}
O_{1}\,=\,(\bar{b}_{\alpha}c_{\beta})_{V-A}(\bar{u}_{\beta}D_{\alpha})_{V-A},
\;O_{2}\,=\,(\bar{b}_{\alpha}c_{\alpha})_{V-A}(\bar{u}_{\beta}D_{\beta})_{V-A}.
\end{eqnarray}
where $\alpha$ and $\beta$ are the color indices,
$(\bar{b}_{\alpha}c_{\beta})_{V-A}\,=\,\bar{b}_{\alpha}\gamma^{\mu}(1-\gamma^{5})c_{\beta}$.
Conventionally, we define the combined Wilson coefficients as
\begin{eqnarray}
a_{1}=C_{2}+C_{1}/3,\;a_{2}=C_{1}+C_{2}/3.
\end{eqnarray}

In the hadronic matrix element calculation, we factorize the decay
amplitude into soft($\Phi$), hard(H), and harder (C) dynamics
characterized by different scales \cite{0511239,prd81014002},
\begin{eqnarray}
\mathcal
{A}\;\sim\;&&\int\,dx_{1}dx_{2}dx_{3}b_{1}db_{1}b_{2}db_{2}b_{3}db_{3}\nonumber\\
&&\times
Tr\left[C(t)\Phi_{B}(x_{1},b_{1})\Phi_{M_{2}}(x_{2},b_{2})\Phi_{M_{3}}(x_{3},b_{3})H(x_{i},b_{i},t)S_{t}(x_{i})e^{-S(t)}\right].
\end{eqnarray}
where  $b_{i}$ is the conjugate variable of quark's transverse
momentum $k_{iT}$, $x_{i}$ is the momentum fractions of valence
quarks,and $t$ is the largest energy scale in function
$H(x_{i},b_{i},t)$ which is the hard part.  $C(t)$ are the Wilson
coefficients with resummation of the large logarithms $\ln(m_{W}/t)$
produced by the QCD corrections of four quark operators.
$S_{t}(x_{i})$ is the jet function, which is obtained by the
threshold resummation and smears the end-point singularities on
$x_{i}$ \cite{prd66094010}. The last term, $e^{-S(t)}$, is the
Sudakov form factor, from resummation of double logarithms, which
suppresses the soft dynamics effectively and the long distance
contributions in the large $b$ region \cite{prd57443,lvepjc23275}.
Thus it makes the perturbative calculation of the hard part $H$
applicable at intermediate scale, i.e., $m_{B}$ scale. The $\Phi_i$,
meson wave functions, are nonperturbative input parameters but
universal for all decay modes.

The lowest
order Feynman diagrams of the considered decays are shown in Fig.1. The amplitude
from factorizable diagrams (a) and (b) in Fig.1 is

\begin{figure}[]
\begin{center}
\vspace{-5cm} \centerline{\epsfxsize=10 cm \epsffile{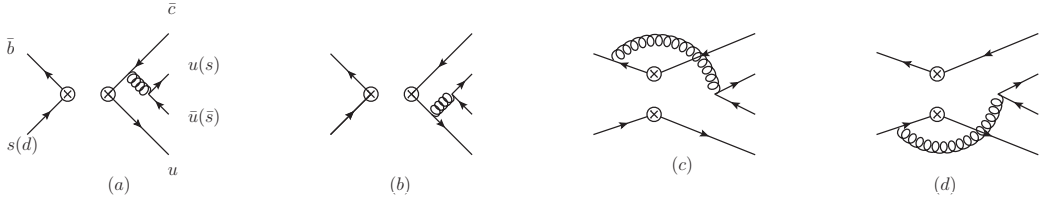}}
\vspace{-6cm} \caption{annihilation diagrams contributing to the
$B\,\rightarrow\,\bar{D}T$ decays in PQCD}
 \label{fig:lodiagram}
 \end{center}
\end{figure}
\begin{eqnarray}
\mathcal {A}_{af}&=&8\sqrt{\frac{2}{3}}C_{F}f_{B}\pi
m_{B}^{4}\int_{0}^{1}\,dx_{2}dx_{3}\int_{0}^{1/\Lambda}\,b_{2}db_{2}b_{3}db_{3}\,\phi_{D}(x_{2},b_{2})\nonumber\\
&&\times\left\{\left[-\phi_{T}(x_{3})x_{2}+2r_{D}r_{T}\phi_{T}^{s}(x_{3})(x_{2}+1)\right]\right.\nonumber\\
&&\left.\cdot h_{af}((1-x_{3}),x_{2}(1-r_{D}^{2}),b_{2},b_{3})E_{af}(t_{e})\right.\nonumber\\
&&\left.-\left[\phi_{T}(x_{3})(x_{3}-1)-r_{D}r_{T}(\phi_{T}^{t}(x_{3})(1-2x_{3})+\phi_{T}^{s}(x_{3})(2x_{3}-3))\right]\right.\nonumber\\
&&\cdot\left.h_{af}(x_{2},(1-x_{3})(1-r_{D}^{2}),b_{3},b_{2})E_{af}(t_{f})\right\}.
\label{af}
\end{eqnarray}
In this function, $C_{F}=4/3$ is the group factor of $SU(3)_{c}$, and $r_{D(T)}=m_{D(T)}/m_{B}$. The hard scale
 $t_{e,f}$ and the functions $E_{af}$ and $h_{af}$ are given by
\begin{eqnarray}
&&t_{e}\,=\,\max\{\sqrt{x_{2}(1-r_{D}^{2})}m_{B},1/b_{2},1/b_{3}\},\nonumber\\
&&t_{f}\,=\,\max\{\sqrt{(1-x_{3})(1-r_{D}^{2})}m_{B},1/b_{2},1/b_{3}\},
\label{taf}
\end{eqnarray}
\begin{eqnarray}
&&E_{af}(t)\,=\,\alpha_{s}(t)\cdot \exp[-S_{T}(t)-S_{D}(t)],
\end{eqnarray}
\begin{eqnarray}
h_{af}(x_{2},x_{3},b_{2},b_{3})\,&=&\,(\frac{i\pi}{2})^{2}H_{0}^{(1)}\left(\sqrt{x_{2}x_{3}}m_{B}b_{2}\right)\nonumber\\
&&\left[\theta(b_{2}-b_{3})H_{0}^{(1)}\left(\sqrt{x_{3}}m_{B}b_{2}\right)J_{0}\left(\sqrt{x_{3}}m_{B}b_{3}\right)\right.\,+\nonumber\\
&&\left.\theta(b_{3}-b_{2})H_{0}^{(1)}\left(\sqrt{x_{3}}m_{B}b_{3}\right)J_{0}\left(\sqrt{x_{3}}m_{B}b_{2}\right)\right]\cdot
S_{t}(x_{3}).
\end{eqnarray}
The amplitude
for nonfactorizable diagrams (c) and (d) in Fig.1 is
\begin{eqnarray}
\mathcal {M}_{anf}&=&\frac{32}{3}C_{F}\pi
m_{B}^{4}\int_{0}^{1}\,dx_{1}dx_{2}dx_{3}\int_{0}^{1/\Lambda}\,b_{1}db_{1}b_{2}db_{2}\,\phi_{B}(x_{1},b_{1})\phi_{D}(x_{2},b_{2})\nonumber\\
&&\times \left\{\left[\phi_{T}(x_{3})x_{2}+r_{D}r_{T}(\phi_{T}^{s}(x_{3})(x_{3}-x_{2}-3)+\phi_{T}^{t}(x_{3})(x_{2}+x_{3}-1))\right]\right.\nonumber\\
&&\left.\cdot h_{anf1}(x_{1},x_{2},x_{3},b_{1},b_{2})E_{anf}(t_{g})\right.\nonumber\\
&&\left.+\left[\phi_{T}(x_{3})(x_{3}-1)+r_{D}r_{T}(\phi_{T}^{s}(x_{3})(x_{2}-x_{3}+1)+\phi_{T}^{t}(x_{3})(x_{2}+x_{3}-1))\right]\right.\nonumber\\
&&\cdot\left.h_{anf2}(x_{1},x_{2},x_{3},b_{1},b_{2})E_{anf}(t_{h})\right\}.
\end{eqnarray}
\begin{eqnarray}
t_{g}\,&=&\,\max\{\sqrt{x_{2}(1-x_{3})(1-r_{D}^{2})}m_{B},\sqrt{1-(1-(1-x_{3})(1-r_{D}^{2}))(1-x_{1}-x_{2})}m_{B},\nonumber\\
&&1/b_{1},1/b_{2}\},\nonumber\\
t_{h}\,&=&\,\max\{\sqrt{x_{2}(1-x_{3})(1-r_{D}^{2})}m_{B},\sqrt{(1-x_{3})(1-r_{D}^{2})|x_{1}-x_{2}|}m_{B},\nonumber\\
&&1/b_{1},1/b_{2}\},
\label{tanf}
\end{eqnarray}
\begin{eqnarray}
E_{anf}\,=\,\alpha_{s}(t)\cdot
\exp[-S_{B}(t)-S_{T}(t)-S_{D}(t)]\mid\,_{b_{2}=b_{3}},
\end{eqnarray}
\begin{eqnarray}
h_{anfj}(x_{1},x_{2},x_{3},b_{1},b_{2})\,&=&\,\frac{i\pi}{2}\left[\theta(b_{1}-b_{2})H_{0}^{(1)}\left(Fm_{B}b_{1}\right)J_{0}\left(Fm_{B}b_{2}\right)\right.\nonumber\\
&&\left.+\theta(b_{2}-b_{1})H_{0}^{(1)}\left(Fm_{B}b_{2}\right)J_{0}\left(Fm_{B}b_{1}\right)\right]\nonumber\\
&&\times \left\{\begin{array}{ll}
\frac{i\pi}{2}H_{0}^{(1)}\left(\sqrt{|F_{j}^{2}|}m_{B}b_{1}\right),&
F_{j}^{2}<0,\\
K_{0}\left(F_{j}m_{B}b_{1}\right),& F_{j}^{2}>0,
\end{array}\right.
\end{eqnarray}
with $j=1,2$.
\begin{eqnarray}
F^{2}&=&x_{2}(1-x_{3})(1-r_{D}^{2}),\nonumber\\
F_{1}^{2}&=&1-(1-(1-x_{3})(1-r_{D}^{2}))(1-x_{1}-x_{2}),\nonumber\\
F_{2}^{2}&=&(1-x_{3})(1-r_{D}^{2})(x_{1}-x_{2}).
\end{eqnarray}
The expressions of $S_{B}(t)$, $S_{T}(t)$, $S_{D}(t)$ and $S_{t}$ can be found in ref.\cite{prd66094010,lvepjc23275,prd63074009, epjc28515}. The wave functions of initial and final states can be found in ref.\cite{plb622,prd76074018,prd71054025,prd78014018,jpg37015002,prd81034006,prd83014008,zheng1,zheng2}.

With the functions obtained in the above, the amplitudes of
 these pure annihilation type decay channels can be given by
\begin{eqnarray}
\mathcal {A}(B^{0}\rightarrow
D_{s}^{-}K_{2}^{*+})=\frac{G_{F}}{\sqrt{2}}V_{cb}^{*}V_{ud}[a_{2}\mathcal
{A}_{af}+C_{2}\mathcal {M}_{anf}],
\label{1}
\end{eqnarray}
\begin{eqnarray}
\mathcal {A}(B_{s}^{0}\rightarrow
\bar{D}^{0}a_{2}^{0})=\frac{G_{F}}{\sqrt{2}}\frac{1}{\sqrt{2}}V_{cb}^{*}V_{us}[a_{2}\mathcal
{A}_{af}+C_{2}\mathcal {M}_{anf}],
\label{2}
\end{eqnarray}
\begin{eqnarray}
\mathcal {A}(B_{s}^{0}\rightarrow
D^{-}a_{2}^{+})=\frac{G_{F}}{\sqrt{2}}V_{cb}^{*}V_{us}[a_{2}\mathcal
{A}_{af}+C_{2}\mathcal {M}_{anf}].
\label{3}
\end{eqnarray}

\section{NUMERICAL RESULTS AND DISCUSSIONS}
For numerical analysis, we use the following input parameters:
\begin{eqnarray}
&&f_{B/B_{s}}=0.21/0.23\,GeV,\,\,f_{D/D_{s}}=0.205/0.241\,GeV,\,f_{K_{2}^{*}}^{(T)}=118(77)\,MeV,\nonumber\\
&&f_{a_{2}}^{(T)}=102(117)\,MeV,\,M_{D/D_{s}}=1.869/1.968\,GeV,\,M_{B/B_{s}}=5.279/5.366\,GeV,\nonumber\\
&&|V_{cb}|=0.0415\pm0.0011,\,|V_{ud}|=0.9742\pm0.0002,\,|V_{us}|=0.2257\pm0.0012,\nonumber\\
&&\Lambda_{QCD}^{f=4}=0.25\,GeV.
\end{eqnarray}
After numerical calculation, the branching ratios of these decays are:
\begin{eqnarray}
&&Br(B^0\rightarrow D_{s}^{-}K_{2}^{*+})=(60.6_{-16.5\,-10.4\,-2.1}^{+17.3\,+4.3\,+3.2})\times 10^{-6},\nonumber\\
&&Br(B_{s}\rightarrow \bar{D}^{0}a_{2}^{0})=(1.1_{-0.4\,-0.2\,-0.1}^{+0.4\,+0.1\,+0.1})\times 10^{-6},\nonumber\\
&&Br(B_{s}\rightarrow D^{-}a_{2}^{+})=(2.3_{-0.8\,-0.4\,-0.1}^{+0.8\,+0.2\,+0.1})\times 10^{-6}.
\end{eqnarray}

The branching ratio obtained from the analytic formulas may be sensitive to many parameters especially those in
the meson wave function. For the theoretical uncertainties in our calculations, we estimated
three kinds of them:
The first errors in our calculations are caused by the hadronic
parameters, such as the decay constants and the shape parameters in
wave functions of charmed meson and the $B_{(s)}$ meson, and the decay constants of tensor
mesons. The second errors are estimated from the unknown
next-to-leading order QCD corrections with respect to $\alpha_{s}$
and nonperturbative power corrections with respect to scales in
Sudakov exponents, characterized by the choice of the
$\Lambda_{QCD}\,=\,(0.25\,\pm\,0.05)$ GeV and the variations of the
factorization scales defined in eq.\ref{taf} and eq.\ref{tanf}. The third error is from
the uncertainties of the CKM matrix elements. It is easy to see that
the most important theoretical uncertainty is caused by the
non-perturbative hadronic parameters, which are universal and can be improved by
experiments.

These pure annihilation type decays considered in this work are
dominant by W exchange diagram. All these decays do not have
contributions from the penguin operators. Since the direct CP
asymmetry is caused by the interference between the contributions of
tree operators and that of penguin operators, it does not appear in
these modes. Although the annihilation type diagrams are power
suppressed in PQCD approach, the branching ratio of these considered
Cabibbo-Kobayashi-Maskawa-favored decays are sizable and large
enough to be measured in experiment. Through the study of these pure
annihilation type decay modes, we can understand the annihilation
mechanism in $B$ physics well.

It is easy to find that there are large theoretical uncertainties in
any of the individual decay mode calculations. However, we can
reduce the uncertainties by ratios of decay channels. For example,
simple relations among these decay channels are derived from
eq.(\ref{1}-\ref{3})
\begin{eqnarray}
&&\frac{Br(B^0\rightarrow D_{s}^{-}K_{2}^{*+})}{Br(B_{s}\rightarrow \bar{D}^{0}a_{2}^{0})}\sim\frac{2f_{D_{s}}^{2}V_{ud}^{2}}{f_{D}^{2}V_{us}^{2}} \sim 60 \sim \frac{60.6}{1.1},\nonumber\\
&&\frac{Br(B_{s}\rightarrow \bar{D}^{0}a_{2}^{0})}{Br(B_{s}\rightarrow D^{-}a_{2}^{+})}\sim\frac{1}{2}\sim \frac{1.1}{2.3}.
\end{eqnarray}
It is obvious that any significant deviation from the above
relations will be a signal of new physics.

\section{SUMMARY}
We calculate the branching ratios of three pure annihilation type
decays in the perturbative QCD approach. The predicted branching
ratios are $Br(B^0\rightarrow D_{s}^{-}K_{2}^{*+})\sim 6\times
10^{-5}$, $Br(B_{s}\rightarrow \bar{D}^{0}a_{2}^{0})\sim 1\times
10^{-6}$ and $Br(B_{s}\rightarrow D^{-}a_{2}^{+})\sim 2 \times
10^{-6}$. They are sizable and large enough to be measured in
forthcoming experiment. The study about the pure annihilation type
decays can help us understand the annihilation mechanism in $B$
physics. There are no direct CP asymmetries, because these decays
have no contributions from penguin operators in the standard model.

$\textbf{Acknowledgment}$

 We are very grateful to Xin Yu and
Dr. Run-Hui Li for helpful discussions. This Work is supported by
the National
Science Foundation of China under the Grant No.11075168.

\end{document}